\documentclass[preprintnumbers,amsmath,amssymb]{revtex4}
\usepackage[]{epsfig}
\newcommand{\xieq}{\xi_{\rm eq}}
\usepackage{graphicx}
\usepackage{dcolumn}
\usepackage{bm}

\begin{document}

\title{Hiking through glassy phases: physics beyond aging}

\author{Ludovic Berthier$^1$, Virgile Viasnoff$^2$, Olivia White$^3$,
Vladimir Orlyanchik$^4$,
and Florent Krzakala$^5$}
\affiliation{
$^1$Theoretical physics, 1 Keble Road, Oxford, OX1 3NP, UK \\
$^2$LPM, ESPCI, 10 rue Vauquelin, 75005 Paris, France \\
$^3$Jefferson Laboratories, Harvard University,
Cambridge, MA 02140 USA\\
$^4$The Racah Institute of Physics, 
The Hebrew University, Jerusalem 91904, Israel\\
$^5$LPTMS, B\^at. 100, Universit\'e Paris-Sud, 91406 Orsay, France}

\date{\today}

\begin{abstract}
Experiments performed on a wide range of glassy materials display many 
interesting phenomena, such as aging behavior. 
In recent years, a large body of experiments probed this 
nonequilibrium glassy dynamics through elaborate protocols, in which
external parameters are shifted, or cycled in the course
of the experiment.
We review here these protocols, as well as 
experimental and numerical results.
Then, we critically discuss various theoretical approaches
put forward in this context.
Emphasis is put more on the generality of the phenomena than
on a specific system. 
Experiments are also suggested.
\end{abstract}

\maketitle

\section{Introduction}
\label{intro}
In this summer school, we were given many examples of glassy systems, 
glassy dynamics, and glass transitions, even though a proper
definition of the word `glassy' was not really provided.
However, all glassy materials share the property that
their relaxation times are extremely large compared to the time
scale of a typical experiment,
at least in a part of their phase diagram.
For practical purposes, they are thus out of equilibrium, meaning
that in principle the whole sample history is relevant to a description 
of their physical properties. 
This paper is dedicated to the study of some specific histories
applied to various glassy materials.

As physicists, we want to study the simplest histories that allow for an 
understanding of all the relevant mechanisms at work.
Hopefully, an understanding of simple protocols will
also allow for prediction or calculation of the behavior 
resulting from increasingly elaborate procedures.

\begin{figure}
\begin{center}
\psfig{file=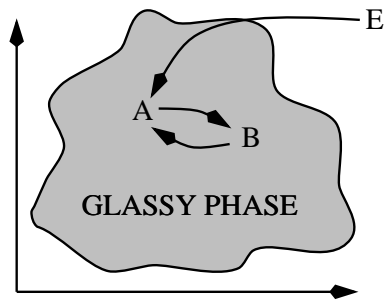,width=5cm}
\end{center}
\caption{Schematic view of the phase diagram of a glassy material. The 
relaxation time of the system in the shaded area is too large for the system
to equilibrate on an experimental time scale. The arrows depict 
simple aging ($E \to A$), shifts ($E \to A \to B$), or
cycling ($E \to A \to B \to A$) experiments.
We do not label the axis, since the paper deals with an Anderson 
insulator, a colloidal suspension, a spin glass model, and a ferromagnet 
at criticality.}
\label{phasediag}
\end{figure}

Over the last decades, the experiment most often performed
has been the {\it simple aging} experiment, see
Fig.~\ref{phasediag}. The system is 
quenched from a non-glassy part of the phase diagram, $E$, into 
the glassy phase, $A$.
The system relaxation time is so large that 
all its physical properties continue to evolve slowly with time (aging).
This phenomenon has been known for a long time  
in the field of structural glasses~\cite{struik}
 before its rediscovery in the field 
of spin glasses~\cite{agingSG}.
Interestingly, aging is observed in a still increasing
number of experimental systems such as soft materials (like 
pastes~\cite{paste},
colloidal suspensions~\cite{luca}, or clays~\cite{lapo}), 
dipolar glasses~\cite{dipo}, disordered ferromagnets~\cite{ferrom} 
and ferroelectrics~\cite{ferroe}, granular matter~\cite{granus}, 
superconductors~\cite{supercon}, etc.
Loss of stationarity is best illustrated by the study of two-time
quantities. 
One typically computes
the correlation between times $t$ and $w$, $C(t,w)$,
or the response of the system at time $t$ to a perturbation applied 
at $w$, or equivalently the 
time evolution of a susceptibility at frequency $\omega$, $\chi(\omega,t)$.
In most common experimental regimes, 
two-time quantities scale as $C(t,w) \sim {\cal F}_C (t/w)$, 
$\chi(\omega,t) \sim {\cal F}_\chi (\omega t)$, where
${\cal F}_\ast(x)$ denotes a scaling function.
This scaling indicates
that after a time $t$ the only relevant time scale
in the system is the time $t$ itself.

In order to probe the dynamics of the glassy 
phase in greater detail,
more elaborate and systematic  
experimental protocols have been performed~\cite{reviewmanip}, 
in which some external parameters are shifted, 
or cycled, during the experiment, see Fig.~\ref{phasediag}.
Such experiments reveal spectacular new phenomena, such
as rejuvenation and memory effects. 
Such new effects must be accounted for by any 
theory of aging, possibly allowing for discrimination between different 
theoretical approaches to glasses and aging phenomena.
In addition, more detailed experiments may allow 
for discrimination between different families of glassy systems, and thus 
may help theoreticians refine their description of specific glassy systems.
A large number of recent experimental papers are dedicated 
to such experiments, on a wide 
variety of systems, making the subject a very active one. By contrast,
these protocols are barely mentioned in the classic
theoretical review in the field~\cite{reviewth}
 and it is one of this paper's purpose
to fill this little gap.

In the first part of the paper we review briefly 
the basic experimental facts.
Next we discuss two mean-field theoretic approaches to the problem.
We then formulate the rudiments of a simple, but fairly robust, 
phenomenology in terms of          
length scales which grow with time and
discuss realizations of this 
scenario.
Last, we show that a nice account of both recent experiments performed
on an Anderson insulator and on a colloidal suspension
can be given in terms of such growing length scales.

\section{Experimental facts}
\label{phenomeno}

In this section, we present the main experimental facts
observed when the protocols of Fig.~\ref{phasediag} are 
actually performed.
This will allow us also to define precisely the vocabulary used
throughout the paper. 
We use our own data to describe these phenomena, but emphasize 
that a similar phenomenology has been observed in many different systems.
Such experiments were first performed on
spin glasses (rejuvenation and memory 
effects~\cite{reviewmanip,cycle1,cycle2,cycle3}) 
and polymer glasses  (mechanical rejuvenation~\cite{mckenna}
and Kovacs effect~\cite{struik,kovacs,reviewappl}) 
in temperature shift or cycling experiments.
Thus, we adopt `temperature' as a control parameter, but also discuss
the case of other control parameters.
The degree to which different such external parameters are equivalent is
a completely open question. It
was recently asked in the present context~\cite{virgile}, 
as a part of a more general 
research line~\cite{jamming,cates}.
Here, we take a pragmatic approach and 
elaborate on experimental similarities.
Sections \ref{rejuvenation}, \ref{underaging}, and \ref{kovacs} deal with 
effects encountered in shift experiments, while section \ref{memory}
deals with cycling experiments.

\subsection{Rejuvenation}
\label{rejuvenation}

\begin{figure}
\begin{center}
\psfig{file=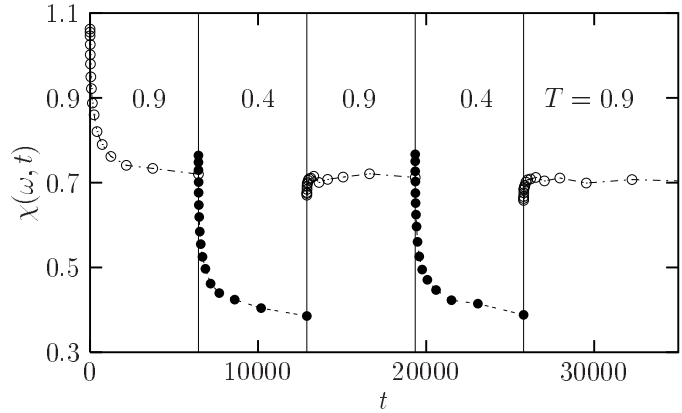,width=9.cm}
\end{center}
\caption{A temperature cycling experiment performed numerically in a spin 
glass model. The system is quenched at $t=0$ in the spin glass phase, 
$T_A<1.0$.
It ages at the temperature $T_A=0.9$ as demonstrated by the time evolution
of a quantity analogous to a magnetic susceptibility.
The temperature is then shifted at $t_A=6450$ 
to $T_B=0.4$ where rejuvenation takes place.
At $t_B=12900$, the temperature is shifted back 
to $T_A=0.9$, demonstrating,
after a very short transient, the memory effect.
A second cycle in then performed.}
\label{effects1}
\end{figure}

First consider a shift experiment, see Fig.~\ref{phasediag}.
The system is quenched 
at initial time $t=0$ from a high temperature to a temperature $T_A$ in 
the glassy phase. For $0< t <t_A$, the temperature is kept 
constant. Up to this point, this is a simple aging experiment, manifested
by the slow evolution 
of physical quantities. 
Such typical slow evolution 
is shown in the left part of Fig.~\ref{effects1}
where a quantity analogous to a magnetic susceptibility at given frequency
is computed in the numerical simulation of a 
microscopic spin glass model~\cite{BB}.

At $t=t_A$, the temperature is shifted to $T_B$.
As seen in Fig.~\ref{effects1}, aging is restarted by a 
negative shift 
($ T_A > T_B $)
in the sense that 
the resulting curve is similar to that obtained in a 
direct quench to $T_B$. This restart of the dynamics is 
called {\it rejuvenation effect} because the time $t_A$ spent 
at $T_A$---the sample `age'---seems 
to have no influence on the dynamics following the shift. 
The same effect is obtained if $T_B > T_A$.
We note also that the term `rejuvenation' was first employed
to describe the effect of large stresses on the aging 
of polymers~\cite{struik,mckenna}.

\subsection{Overaging and underaging}
\label{underaging}

\begin{figure}
\begin{center}
\begin{tabular}{cc}
\psfig{file=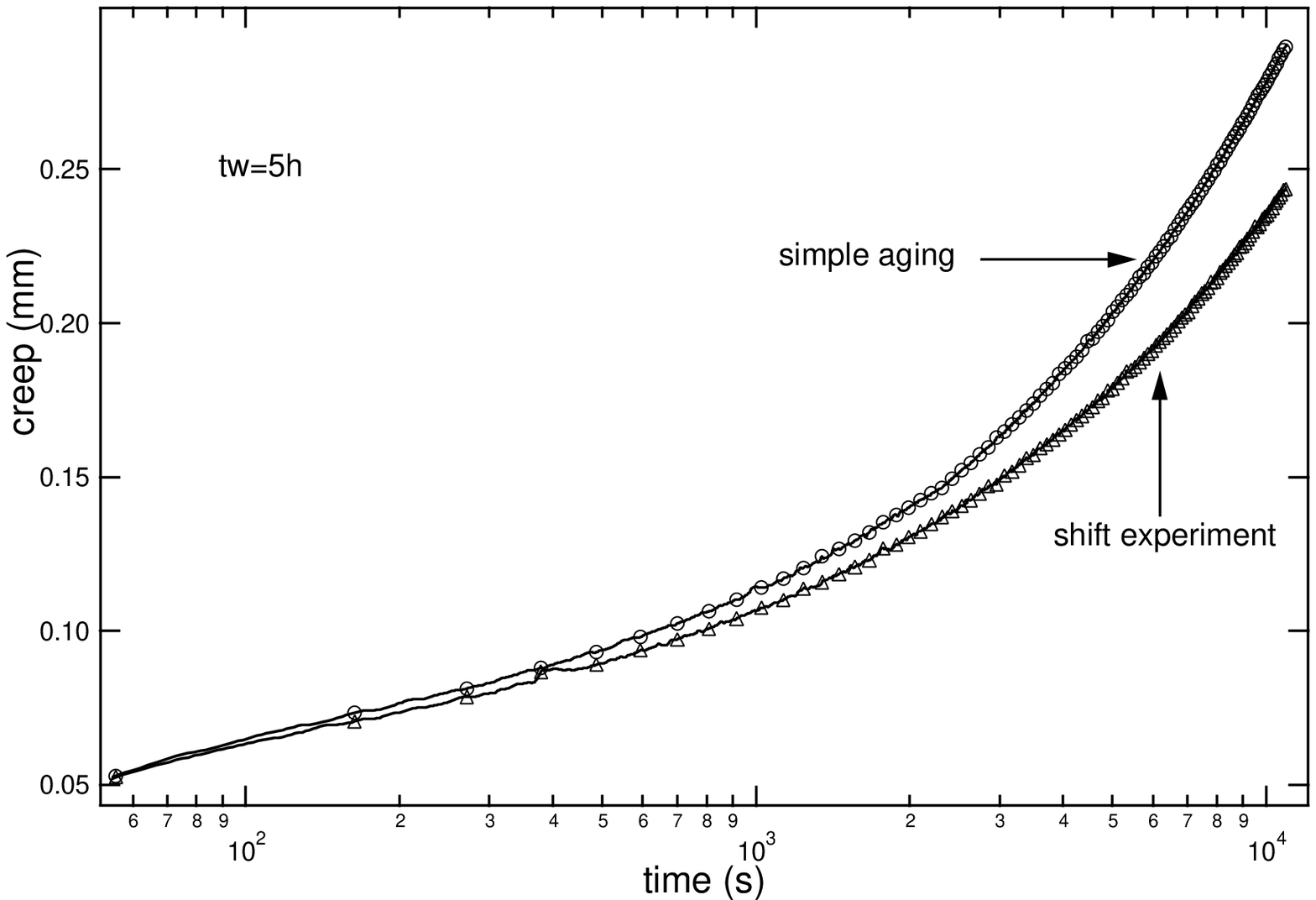,width=7.5cm} &
\psfig{file=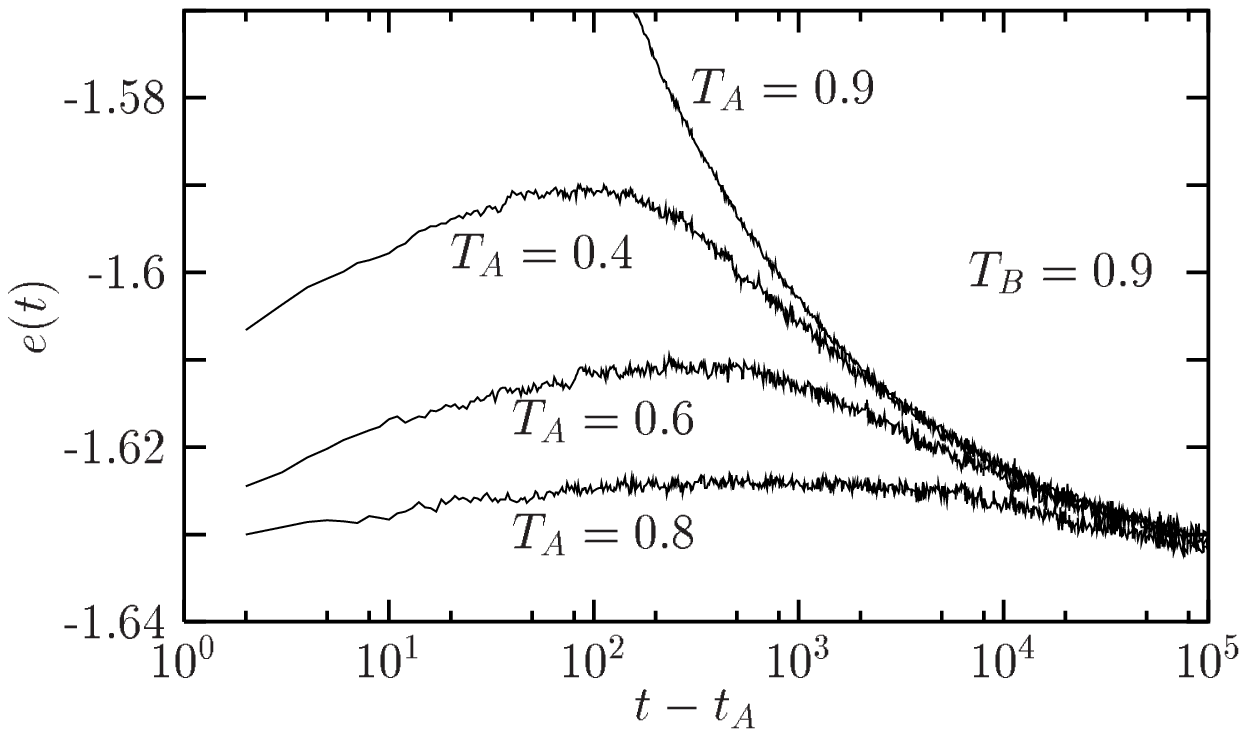,width=7.5cm} 
\end{tabular}
\end{center}
\caption{Left: overaging observed in a shift experiment with $T_B < T_A$. 
The creep compliance of a polymer after the shift (triangles)
is compared to the curve obtained in a simple aging experiment at $T_B$
(circles) and the same aging time, $t_A=5$h.
The long-time response is clearly slower after the shift.
Right: Kovacs effect observed in a shift experiment with $T_B > T_A$. 
The energy density of a spin glass model recorded after the shift
exhibits a typical nonmonotonic behavior.}
\label{effects2}
\end{figure}

Obviously, if $T_B = T_A$, no rejuvenation takes place, meaning 
that 
in order for rejuvenation to be observed in a shift experiment,
$|T_A - T_B|$ must be `large enough'.
For small or intermediate 
$|T_A - T_B|$, 
a phenomenon recently called 
{\it overaging} is observed~\cite{virgile}, when $T_A > T_B$.
This is illustrated in Fig.~\ref{effects2} (left),
where results obtained with a polymer are presented~\cite{virgile2}.
In this figure, two-time linear response to a stress 
step (creep compliance) after a shift 
from $ T_A $ to $ T_B $
is compared to that obtained in a simple aging 
experiment 
at $ T_B$ 
with the same aging time. 
The response after the shift is slower than 
that obtained in the 
the simple aging experiment, and the system looks `older', 
or `overaged'.
Similarly, an {\it underaging} would be obtained 
if $T_A < T_B$.

These effects were also observed experimentally in a colloidal
suspension submitted to a transient oscillatory shear~\cite{virgile}, 
as well as in temperature shift protocols in 
experimental~\cite{reviewmanip,cycle2}
and numerical~\cite{BB,yosh} studies of spin glasses.

\subsection{Memory effect of the first kind, or `Kovacs effect'}
\label{kovacs}

A {\it memory effect} takes place in the shift protocol 
when $T_B > T_A$. It was first observed
by Kovacs in polymers~\cite{kovacs}.
To distinguish it from a second memory effect (see below), this effect
was called {\it Kovacs effect} in Ref.~\cite{surf}.
Kovacs measured the specific volume, $V(t)$, 
of the polymer during the experiment, but
other physical quantities (index of refraction, energy density,...) can 
be investigated. 
The effect is particularly striking when $t_A$ is chosen 
so that $V(t_A) = V_{\rm eq}(T_B)$, 
which means that immediately
after the quench, the volume has already reached its equilibrium value
at the new temperature. Hence, a naive expectation would be 
that $V(t>t_A) = const = V_{\rm eq}(T_B)$. Instead Kovacs observed
a nonmonotonic variation of the volume showing that the system has some 
{\it memory} of its state at the initial temperature.
This experiment is reproduced 
in Fig.~\ref{effects2} using the energy density of 
a microscopic spin glass model
in a numerical simulation~\cite{BB}.
Similar results have been obtained in supercooled liquids~\cite{leheny}, 
granular materials~\cite{granus}, foams~\cite{foam} 
or dipolar glasses~\cite{dipo}.

\subsection{Memory effect of the second kind}
\label{memory}
Let us describe the continuation of the experiment shown in  
Fig.~\ref{effects1}. At time $t_B > t_A$, $T$
is shifted backed to its initial value $T_A$. 
After a very short transient, $\chi(\omega,t)$ resumes its evolution as
if there had been no aging at $T_B$. The system has a
{\it memory} of the first stage, despite
the strong rejuvenation observed in the intermediate stage of the cycle.
Simultaneous observation of rejuvenation 
and memory is spectacularly demonstrated in the `dip 
experiment'~\cite{cycle3}.
The protocol is essentially a cycling experiment in which the temperature 
is decreased at a fixed, finite rate (instead of at an infinite
rate as in an idealized cycle).
The ramp from high temperature stops at $T_A$, where the temperature 
is kept constant a time $t_A$ after which
cooling is resumed.
In this context 
rejuvenation means that the further evolution of the system is almost
the same with or without the stop at $T_A$. 
The temperature is then raised back at
a constant rate. 
Memory means that near $T_A$, the system `remembers' its 
stop~\cite{cycle3,ludovic}
and behaves differently from a system not held at $T_A$. 
However, this experiment 
is less simple to analyze theoretically because it mixes
cooling rate effects with rejuvenation and memory.
For this reason, we stick to cycles in 
what follows. 

\subsection{Need for a generic and robust phenomenology}

This quick experimental
review shows that a number of effects are 
both 
highly non-trivial 
as well as 
generically observed in a wide 
range of materials, suggesting that these phenomena 
are intrinsic to nonequilibrium glassy dynamics. 
This has two immediate consequences. 
(1) Any phenomenological theory of aging must 
account for these effects 
in addition to
the results of simple aging experiments. 
(2) The phenomenology should be based upon general considerations,
which themselves proscribe the range of experimental systems to which 
the theory applies. Ideally this would be the large range of quite 
different experimental systems in which glassy dynamics is observed. 

\section{Two mean-field theoretical approaches}

As reviewed in Ref.~\cite{reviewth}, 
theoretical approaches to aging phenomena can be
classified in three large families. We analyze 
two mean-field ones in this subsection in view of 
the above experimental facts.

\subsection{Trap and multi-trap models}

A first approach, made popular through the `trap model' formulated
in Ref.~\cite{trap}, 
considers the dynamics from the point of view of 
phase space and describes the dynamics of a  
point particle evolving through a given energy landscape.
The reduction of a many-body system to a one-particle problem
classifies this approach
in the mean-field family.
The `complexity' stems from an assumed large distribution for the free energy  
of the metastable states,
typically chosen to be exponential,
$\rho(E) \sim \exp(-E/T_g)$.
If one further assumes that energy barriers are related to the free energy of 
the states, one gets the
distribution of trapping times
$\rho(\tau) \sim \tau^{-(1+T/Tg)}$, the first moment of which 
diverges for $T<T_g$. 
The absence of a mean trapping time 
then results 
in the typical $t/w$ scaling of two-time quantities.
Obtaining this scaling in a one-body problem is a remarkable
result.
The crucial drawback of the approach is, obviously, the lack of a 
precise interpretation of what `traps' are. 

To account for the effects described above, 
the model was phenomenologically extended to 
multi-trap models~\cite{multitrap1}, 
each level having its own glass temperature $T_g$.
The phenomenology of cycles is straightforward.
When the temperature is lowered,
new levels of the hierarchy start to age (rejuvenation effect), while higher 
levels are completely frozen and unaffected by the stay at low 
temperature (memory).
This `hierarchical phase space picture' has often been invoked 
by the Saclay group to interpret experiments on spin 
glasses~\cite{cycle2}.
Furthermore, numerical simulations of concrete realizations 
of multi-trap models confirm that this picture 
satisfactorily  reproduces
the experiments~\cite{multitrap2}.  
As a similar realization of this picture, the Sinai model was 
recently studied in this context~\cite{sinai}.

\subsection{Infinite-range models}
\label{MF}

A more microscopic approach can also be used to describe aging~\cite{reviewth}.
The aim is to solve exactly the 
dynamics starting from the Hamiltonian
of a glassy system. This ambitious program
has been successful for systems with infinite-range interactions.
They are thus mean-field realizations of realistic systems.
Simple aging experiments can be accounted for, 
as described in detail in Cugliandolo's lectures~\cite{leticia}.
These solvable models have proved to be extremely rich,
but by construction they are incomplete descriptions
since real space is completely ignored.

Two families of models 
have 
emerged.
Schematically, the first  
has the phenomenology of structural 
glasses---an example is the $p$-spin model---and
typically exhibits $t/w$ scalings~\cite{cuku37}.
The second family is closer to spin glasses---the prototype being
the Sherrington-Kirkpatrick model. The behavior of the latter in simple 
aging experiments is quite involved, since it displays
`dynamic ultrametricity'~\cite{cuku38}. 
Without giving the details~\cite{leticia}, 
this implies a complex scaling of two-time functions,
with the presence of a continuous hierarchy of diverging time scales. 
When submitted to shifts and cycles, this second family
(but not the first one)
was shown to exhibit 
rejuvenation and memory asymptotically~\cite{leticia2}.
It would be useful to have simulations of 
these models in order to go
beyond the asymptotic analysis of Ref.~\cite{leticia2}
and 
to get some comparison to experiment.
Within this approach, the glassy effects of the previous section
are explained 
through the existence
of a hierarchy of time scales. 

However, as honestly noted in the conclusion of Ref.~\cite{leticia2}, 
a strong drawback is that the dynamic ultrametricity on which the 
whole interpretation relies is incompatible with experiments.
Obviously this
weakens the general validity of these results.

\section{Spatial approaches}

\subsection{Domain growth}
\label{coarsening}

A third family of models 
describing slow dynamics 
focuses directly on 
spatial aspects~\cite{reviewth}.
Simple aging experiments are 
explained with reference to a characteristic length scale, $\ell(t)$,
which grows with time. 
The physical content of $\ell(t)$ is that 
on scales smaller than $\ell$ the system appears equilibrated while
on larger scales it does not.

A pure Ising ferromagnet quenched from its paramagnetic to its 
ferromagnetic phase provides a simple example of coarsening dynamics.
In the low temperature phase, the ferromagnet has two equilibrium states,
with magnetizations $+$ and $-$.  As time passes, spatial domains
of these equilibrium phases develop and coarsen, with a typical associated 
length scale $\ell(t)$. This domain growth
is driven by surface tension of domain walls.
The dynamic scaling hypothesis is that the only length scale 
involved in this process is the domain size 
$\ell(t)$ itself.  
In a pure ferromagnet, the growth law is temperature independent~\cite{Alan},
$\ell(t) \sim t^{1/2}$,
and this leads to $t/w$ scaling of two-time correlators. 

This phenomenology does not provide
for the rejuvenation and memory effects discussed above and indeed
such effects are not observed in ferromagnets.
For instance, consider rejuvenation.
Thermal reequilibration within the domains is quasi-instantaneous
since thermal fluctuations in the ferromagnetic 
phase are short-range, of typical size $\xieq(T) \ll \ell(t)$. 
Also, since lowering temperature during domain
growth leaves the growth rate unchanged, after a temperature cycle
$T_A \rightarrow T_B \rightarrow T_A$, there cannot be regions
initially equilibrated at $T_A$ that remain unchanged 
during the cycle.
 
However, coarsening in a disordered system cannot be analogous to that
in a pure system~\cite{nopure}.
Coarsened domains will no longer be simple convex objects with 
energy decreasing monotonically with curvature.  
This can be seen readily by considering a 
ferromagnet with some added impurities.
Walls separating $+$ and $-$ domains will tend to avoid 
unusually strong bonds 
and will tend
to be pinned at unusually weak bonds.
In such a disordered system, domain growth will be slower
than it is in a pure ferromagnet.  In the case of high 
disorder, domain growth is likely activated, leading to a 
characteristic domain size $\ell(t) \sim (\ln t)^p$~\cite{RFIM}.
Secondly, in important contrast to pure systems,
renormalization of a disordered system's Hamiltonian
yields statistically similar Hamiltonians.
In particular this means that we should not expect that  the equilibrium
states in the low temperature phase of a disordered system at 
temperatures $T_A$ and $T_B$ are related in any simple way.  
A simple example is given by the effective interaction 
between two spins on the opposite vertices of a square.
Just as the sign of the effective interaction between these two 
spins can change with a perturbation in bond strength, so too 
can it change with a perturbation in temperature~\cite{vincent}.

\subsection{A minimal phenomenology}
\label{mini}

These differences stated, one can ask whether there is a phenomenological
coarsening theory for aging in glassy materials 
analogous to that in systems like pure ferromagnets.
We formulate such 
a `minimal' phenomenology in terms of a growing length scale.  
It must contain the following features.

{\it (i) Existence of a growing coherence length $\ell$}. 
This coherence length has the same physical 
content as described above: objects smaller 
than $\ell$ are quasi-equilibrated while larger objects still 
retain their nonequilibrium initial conditions.
A more precise definition of the `objects' is not necessary at this point.
For clarity we will draw compact domains in our cartoons. 
Furthermore, the coherence length must grow slowly with time
so that equilibrium domains of size $\ell$ remain fixed on
time scales associated with objects smaller than $\ell$.
This first assumption is very natural for some systems, e.g.,
disordered ferromagnets.
Yet for other systems it is very non-trivial. 
For instance, no typical length scale has been invoked yet 
to explain the aging of supercooled liquids or colloidal suspensions.

{\it (ii) Sensitivity of equilibrium on all length scales.}
Obviously, the specific equilibrium state of the system 
depends upon the particular values of the control parameters, but
sensitivity is required in order to provide for the rejuvenation effects.
Of course, even a simple two-level system has to readapt its 
Boltzmann weights upon a temperature change~\cite{JPPRB}.
There remains to be understood what `levels' are 
in a realistic system, and how those evolve with temperature.
In fact as discussed above, sensitivity of the equilibrium state
to control parameter values is a natural consequence of a renormalization 
procedure in a disordered system.  
Note that there are nevertheless questions outstanding. In particular,
which if any glassy systems be described in terms of a glassy phase 
fixed point? 
And even if some can be, is the sort of `chaos' that one gets automatically
in fact responsible for the observed rejuvenation and memory 
phenomena in glassy materials? 
We will return to these points shortly.

{\it (iii) Separation of length scales.}
The length at which a system is equilibrated after a given wait time 
depends strongly on external parameters.  Memory effects are a 
consequence of such dependence.  
This is very natural for the thermally 
activated domain growth expected 
in disordered systems~\cite{JPSTA}:
\begin{equation}
t(\ell,T,\cdots) \sim t_0 \exp \left( \frac{E(\ell,T,\cdots)}{k_B T}
\right),
\label{equation}
\end{equation}
where $t$ 
is the time needed to equilibrate the length scale $\ell$,
$t_0$ a microscopic time scale, and 
$E(\ell, T,\cdots)$ an activation energy that in principle
can depend on the coherence length itself and on temperature.
The dots stand for possible control parameters in addition to $T$.
If $E(\ell) \propto \ell/T$, 
the growth law is logarithmic and 
`super-Arrhenius': $\ell \propto T^2 \ln (t/t_0)$.
If instead barriers grow logarithmically, $E = k_B T_0 
\ln \ell$, one gets 
a $T$-dependent power law growth, $\ell \sim t^{T/T_0}$. 
In either case, thermal activation (Eq.~(\ref{equation})) 
implies that the range of length scales `active' 
in a given time window depends strongly on $T$ and, more generally, 
on other external parameters in systems with glassy 
dynamics~\cite{JPSTA,JPPRB}. 

\subsection{Back to experiments}
\label{back}

\begin{figure}[t]
\begin{center}
\psfig{file=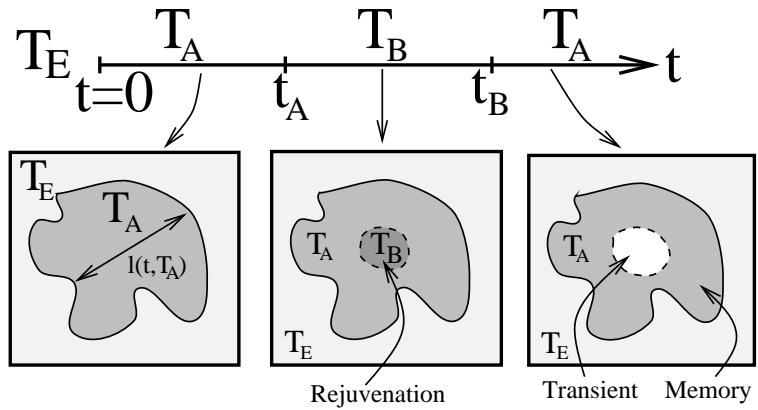,width=10.cm}
\end{center}
\caption{Cartoon suggested by 
the considerations of section \ref{mini} for
a temperature cycling experiment. The related effects are discussed
in section \ref{back}.}
\label{cartoon}
\end{figure}

Consider the cycling experiment $E \to A \to B \to A$ 
of Fig.~\ref{phasediag}.
Again we write the control parameter as a temperature $T$.
The three above assumptions lead to the cartoon of Fig.~\ref{cartoon}
which we now explain.

For times $0<t<t_A$, this is a simple aging experiment
at $T_A$.
Feature (i) entails that
the system equilibrates 
up to a growing coherence length, $\ell(t,T_A)$, leading
to a $\ell(t,T_A) / \ell(w,T_A)$ scaling of two-time functions. This is 
the usual domain growth picture of aging.
Small length scales, $\ell_s < \ell(t,T_A)$, with characteristic
times  $ t_s \ll t $ have reached
equilibrium at $T_A$, while big ones, 
$\ell_b > \ell(t,T_A)$ with characteristic times
$ t_b \gg t $  
still retain their initial nonequilibrium state at $T_E$, 
see Fig.~\ref{cartoon}.

Then the temperature is shifted to $T_B < T_A$ at time $t_A$
with $ |T_A - T_B| $ `large'.
Feature (ii) entails that all length scales,
$\ell \lessgtr \ell(t_A,T_A)$,  
must adapt to the new temperature $T_B$.
This  occurs through the
growth of a new coherence length, $\ell(t-t_A,T_B)$, 
as shown by a dashed line in Fig.~\ref{cartoon}.
Aging is thus restarted from scratch
and {\it rejuvenation due to the reequilibration 
of small length scales} is observed.

At time $t_B$, the system is shifted back to $A$.
Due to (iii), $T_B < T_A$ implies a much slower growth
of $\ell$ during the intermediate stage.
Hence, for $t_B - t_A \approx t_A$, 
as is often the case in experiments, one has 
$\ell(t_B-t_A,T_B) < \ell(t_A,T_A)$ and
the behavior for $t>t_B$ is due to
three types of length scales.
Small length scales, $\ell < \ell(t_B-t_A,T_B)$, have to reequilibrate
at $T_A$. 
Intermediate lengths, $\ell(t_B-t_A,T_B) < \ell <  \ell(t_A,T_A)$,
are already equilibrated at $T_A$. {\it It is in these length scales that
the memory resides}.
Large lengths, $\ell(t_A,T_A) < \ell$, still have to equilibrate at 
$T_A$. 
Hence, after a short transient of duration $\tau$ given
by $\ell(\tau,T_A) \sim \ell(t_B-t_A,T_B) < \ell(t_A,T_A)$, aging
proceeds as the continuation of the first stage.
Note that under the assumption of activated dynamics, simple
inequality of length scales corresponds to a large degree of
inequality in terms of time scales.

If one chooses instead $T_B > T_A$, the
essential situation should be unchanged, provided
that the condition $\ell(t_B-t_A,T_B) < \ell(t_A,T_A)$ is fulfilled,
see Fig.~\ref{cartoon}.
This property is called `symmetrical effect' in the field 
of spin glasses~\cite{reviewmanip}. 
Note, however, that 
this condition on the lengths
is extremely difficult
to satisfy due to 
feature (iii).
Therefore, in most experiments with $T_B > T_A$ the equilibrated length grown 
at $T_B$ is larger than that grown at $T_A$. 
All memory is subsequently erased, resulting in an apparent asymmetry
between positive and negative cycles.

Let us finally turn to Kovacs' experiments.
Two types of length scales must be 
distinguished in order to understand this effect.
Small scales, $\ell < \ell(t_A,T_A)$, are equilibrated at $T_A$
but must adapt to a new, {\it higher}, temperature $T_B$. 
Large lengths, $\ell(t_A,T_A) < \ell$, in contrast,
are still in their nonequilibrium high temperature initial state and must
adapt to a {\it smaller} temperature $T_B$. 
If the energy density $e(t)$ is computed, as in Fig.~\ref{effects2} (right),
small length scales contribute to increase $e(t)$, while large
length scales contribute to decrease $e(t)$. This accounts in
simple terms for the Kovacs effect.

Schematic discussions based on (i), (ii), and (iii) like that
above allow for qualitative prediction of the outcome of given 
experimental protocols.  However, while (i), (ii) and (iii) are natural 
features for a phenomenological theory of coarsening in a glassy
system, in the above they are put in by hand.  Consequently 
we now discuss models that explicitly realize these properties. 

\subsection{Droplets and chaos in spin glasses}

The scaling, or droplet, model of spin glasses~\cite{BM,drop}
possesses these three features and provides quantitative, experimentally 
testable predictions.  These predictions are clearest in the case of
Ising spin glasses, which we discuss here.  
At a given temperature in the spin glass phase, the assumptions 
made by the scaling picture resemble those made in the case of 
pure ferromagnetic systems, but differ due to the disorder
in a spin glass.

The initial assumption of the droplet model is that Ising spin glasses 
have two equilibrium states, related by spin flip symmetry.
Spin glass dynamics are then described in terms of `droplets', low lying 
excitations about these states~\cite{drop}. 
Because of disorder, the boundaries of low lying excitations 
wander in order to take advantage of bonds not satisfied in the 
ground state and to avoid satisfied bonds that are particularly strong.  
Thus droplets are expected to be non-convex with fractal surface 
of dimension $d_f > d - 1$, where $d$ is the space dimensionality.
The droplet model assumes that the energy of  
larger droplets is, on average, larger than that of smaller ones
and hence that the spin glass phase is not destroyed by arbitrarily large,
arbitrarily low energy, excitations.
In particular, it makes the scaling ansatz that the average
free energies of these low lying excitations scales with their size,
$F_{\ell}  \sim \Upsilon(T) \ell^\theta$, $\theta > 0$.
As a result of disorder, $\theta < (d - 1)/2$,
and the distribution of $ F_{\ell} $ is expected to be 
broad, with weight down to zero for all $\ell$.
Further the droplet model assumes that dynamics is activated 
and makes the second scaling assumption that activation 
barriers scale with droplet size, 
$B_\ell \sim \Delta(T) \ell^\psi$ where the distribution of
$B_\ell$ also has weight down to zero.
Stiffness modulii $\Upsilon(T)$ and $\Delta(T)$ go to zero at the critical
temperature $T_c$ as $(1 - T/T_c)^{-\nu \psi}$,
where $\nu$ is the standard exponent.

Feature (i) given above follows immediately from the assumptions of
the scaling picture. Aging proceeds by equilibration of droplets in 
increasingly larger equilibrium domains of size
$\ell(t,T) \sim (\frac{T}{\Delta(T)}\ln t)^{1/\psi}$.
Feature (iii) also follows immediately both from
activated dynamics and from the temperature dependent 
stiffness coefficients. Thus the droplet
picture predicts simple aging behavior as well as memory effects, 
at least under certain experimental conditions.

The dynamics provided by the droplet model of spin glasses differs
importantly from the dynamics of a pure ferromagnet 
in the role of temperature change~\cite{drop}.
A spin glass
at $0 < T < T_c $ is indeed described in terms of a renormalized 
Hamiltonian with a renormalized ground state.  However, as mentioned 
above, this renormalized finite temperature Hamiltonian is only 
statistically similar to the zero temperature Hamiltonian. 
Hence the equilibrium states and droplet excitations thereof at various 
temperatures are not related to one another in any simple way.  
In fact, a straightforward statistical argument~\cite{drop,chaosBM}
provides an overlap length associated with a small temperature
change from $T$ to $T + \Delta T$, given by 
$\ell_o(T,\Delta T) \propto |\Delta T|^{-\zeta}$
where $\zeta = \frac{2}{d_f - 2 \theta}$.
The equilibrium state at $T$ is unchanged on scales 
$\ell \ll \ell_o(T,\Delta T)$ but altered significantly on scales
$\ell \gg \ell_o(T,\Delta T)$.
This `temperature chaos' realizes feature (ii) and predicts rejuvenation
when $\ell_o(T, \Delta T)$ is sufficiently small
compared to $\ell(t,T)$~\cite{toydrop}.

Remark that the droplet picture makes certain reasonable, but as yet 
unverified, assumptions.  It assumes that there are equilibrium states
with reference to which low lying excitations can be defined, even
as equilibrium domains are growing.  Such assumptions are still a 
matter for debate~\cite{KM}.
Furthermore, the properties of these excitations are 
given by reasonable but assumed scaling laws. For work
examining the validity these assumptions see 
Refs.~\cite{lamarcq}.
 
Even if the assumptions made in the droplet model are valid, it is
not clear whether this model is sufficient to explain observed 
rejuvenation and memory effects.  
In order for it to do so, the overlap length $\ell_o(T)$
must be small enough to affect observations corresponding to 
experimental length scales.  
We note that the characterization of $\ell_o(T)$  in spin glasses 
is still a very active topic at the theoretical level~\cite{ello}.
A cautious conclusion is that, if $\ell_o$ exists, it certainly
is too large to be observed in numerical simulations of systems like
spin glasses. However,
because $F_\ell$ and $B_\ell$ are both broadly distributed, 
the droplet model anticipates that in rare regions, reorganization 
with change of temperature will take place even on unusually
small length scales and experimentally viable time scales.  
If such anomalous reorganization is common enough,
rejuvenation and memory may still be accounted for by the droplet
model even if the characteristic overlap length $\ell_o(\Delta T)$ itself 
is not readily obtainable in experimental wait times~\cite{reply_yosh}.  
The distribution of length scales about the overlap length
$\ell_o(\Delta T)$ certainly 
requires further theoretical investigation.

On an experimental level, rejuvenation effects
do not constitute a proof of temperature chaos.  
They are simply consistent with its existence. 
In fact, in the simulational study of 
Ref.~\cite{BB} it is argued that a rejuvenation effect is observed in 
a regime in which no chaotic effect of the kind described above
can be detected~\cite{comment}.
Hence, it is worth considering 
other mechanisms that might provide for rejuvenation and memory effects.

\subsection{Surfing on a critical line}
\label{XY}

Another mechanism has recently been invoked to account
for rejuvenation effects, while retaining the simplicity
of the domain growth approach~\cite{surf}.
We showed in section~\ref{coarsening} that
the absence of rejuvenation in standard coarsening
comes from the short-range character of thermal fluctuations.
Since the inequality $\xieq (T) \ll \ell(t)$ prevents rejuvenation, 
the idea of Ref.~\cite{surf} is simply to consider opposite
situations where $\ell(t) \ll \xieq(T)$.
Away from critical points, $\xieq(T)$ is
usually small. The suggestion
is hence to `surf on a critical 
line'~\cite{hike} where $\xieq(T) = \infty$.

A simple aging experiments corresponds then
to a quench to a critical point. 
In this case, the coherence length is equal to the dynamic
correlation length $\ell(t) = \xi(t) \sim t^{1/z_c}$, where $z_c$
is the dynamic critical exponent.
Aging is thus nothing but the successive equilibration
of the critical fluctuations. 
Similarities between this situation and aging in glassy
materials were noted~\cite{GL}.

When the temperature is shifted,
all the critical fluctuations
have to adapt to a new critical point with a new set of critical
exponents. This is the origin
of the rejuvenation effect in this context. 
It is now obvious that all the discussion of section~\ref{back} 
applies, with the dictionary ``objects'' = ``critical 
fluctuations'' and ``coherence length'' = ``dynamic correlation
length''. The mechanism for rejuvenation is 
subtlely different from temperature chaos, even taken close
to the critical point~\cite{nifle}.
Here, all length scales
are always reshuffled without any `overlap length'~\cite{BB,surf}.

\begin{figure}[t]
\begin{center}
\psfig{file=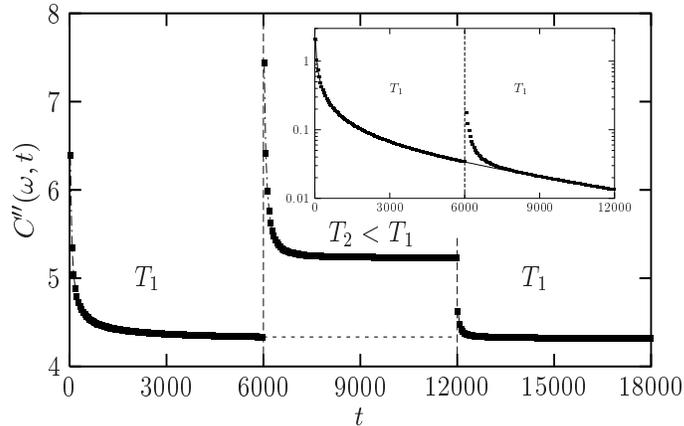,width=9cm}
\end{center}
\caption{Fourier transform
of the spin autocorrelation function of the 2D XY model 
computed analytically in a 
temperature cycling experiment.}
\label{surfing}
\end{figure}

The XY model is a microscopic realization of this picture.
It undergoes, in two dimensions, a Kosterlitz-Thouless transition
from a standard paramagnetic state to a
non-magnetized state with
power-law correlations.
It presents then a `line of critical points', with a 
continuous variation of the critical exponents with temperature.
As a consequence, the 2D XY model exhibits aging, rejuvenation, memory
and Kovacs effects~\cite{surf}. 
The example of a temperature cycle,
where rejuvenation and memory are present,
is given in Fig.~\ref{surfing}.
Note that property (ii) is not naturally present in the model, 
since $\ell(t) = \sqrt{\rho(T) t}$
where $\rho(T)$ is the (renormalized) spin wave stiffness.
To build Fig.~\ref{surfing}, 
$\rho(T)$
was arbitrarily, but not unreasonably, chosen
to ensure the separation of length scales,
$\rho(T_A) = 20 \rho(T_B)$.

This analogy raises the question: are
glassy systems critical? 
A possible answer is yes they are, in which case
they should have in common a line of critical points
below the glass transition, in analogy with the 2D XY model. 
A microscopic mechanism giving rise to such a behavior in, say,
spin glasses, is lacking.
Another is no; but they do experimentally 
behave as if there is a critical line.
The second, more pragmatic
solution requires that $\ell(t)$ is
never decoupled from $\xieq(T)$, even when the latter is finite.
One might argue this if the excitations on length scales
exceeding $\xieq(T)$ are dominated by activation, 
and hence give exponentially increasing times for $\ell(t) \ge \xieq$.
These conditions may make it virtually impossible to enter
the regime $\ell(t) > \xieq(T)$.
Slow evolution in the crossover region could result in effective 
critical behavior with continuous evolution of the exponents.
This is at least fully consistent with results in spin glasses~\cite{BB}.

\section{Two experiments}

\subsection{Anderson insulator}

We mention the Anderson insulator, or `electron glass', 
in this paper since its experimental
properties were discussed at length in this school by 
Ovadyahu~\cite{ova1,ova2}.
Hence, we do not present the system and directly discuss the results.
Also, we only focus here on one of 
the many measurements performed on this system, 
and refer to Refs.~\cite{ova2} for more details.
Our aims are (1) to rephrase the ``aging experiment''
described by Ovadyahu in the present context;
(2) to show that the phenomenology of section~\ref{mini}
might be useful to interpret the data.

The electron glass is first prepared in its `glassy phase'.
This corresponds
to a given low temperature and a gate voltage $V=V_A$, used
as the `control parameter'~\cite{ova1}. This
corresponds then to a `simple aging experiment'.
Aging manifests itself through a logarithmic time evolution of the 
conductance $G$, used as a probe of the dynamics.
Presumably, two-time functions 
would display in this regime interesting scaling behaviors, but
no such measurements have been performed yet.

After a very large time, typically days, $G$ is almost constant.
The gate voltage
is then shifted to $V_B$ during a time $t_w$.
Experiments show that $G$ restarts to evolve 
in a logarithmic way.
This corresponds to a `rejuvenation effect', since the
days already spent in the glassy phase are apparently forgotten.
In a third stage of the experiment, $V$ is shifted back to its initial
value $V_A$. The experiments show that, after a time
of order $t_w$, 
$G$ reaches its initial value again.
This is analogous to what we called the `transient', followed by the
`memory effect of the second kind'.
Furthermore, experiments have shown that during the third stage 
of the experiment (`transient' and `memory') the conductance satisfies
the scaling law $G  = {\cal G}(t/t_w)$ where $t$ is the time
counted from the beginning of the third stage~\cite{ova1,ova2}.
This behavior was called `aging' in Ref.~\cite{ova2}, 
because of the $t/t_w$ scaling,
although the experiment is in fact a complete cycle in terms of $V$.

Can one go beyond words and analogies?
If the similarity between these experiments
and standard temperature cycling experiments
is assumed, then the cartoon of Fig.~\ref{cartoon} 
can be used. 
The $t/t_w$ scaling of the conductance tells that the duration
of the `transient' has the same magnitude as the duration
of the second stage. 
This means that there is almost no
separation of time scales in the electron glass for this range of gate
voltages, as noted in Ref.~\cite{ova2}. 
Building further on analogies, one can reproduce this experiment 
using the 2D XY model. Stay first
a very long (infinite) time at temperature $T_A$, then do a shift of duration
$t_w$ at $T_B$ and come back at $T_A$ at time $t=0$. 
In this case, one can compute the time dependence of
the spin autocorrelation function between times $t$ and $0$, 
noted $C_{t_w}(t,0)$, 
as a function of the duration of the shift
$t_w$. One finds, 
dropping irrelevant factors 
and using the notations of 
section~\ref{XY}:
\begin{equation}
C_{t_w}(t,0) = C_{\rm eq}(t)
\left( \frac{1}{\ell_{T_A}(t)} 
\right)^{\frac{\eta_B - \eta_A}{2}}
\left(
\frac{\Big( 2 \ell_{T_B}^2(t_w) + \ell_{T_A}^2(t) \Big)^2}
{4 \ell_{T_B}^2(t_w)\Big(\ell_{T_A}^2(t)+\ell_{T_B}^2(t_w)\Big)} 
\right)^{\frac{\eta_B - \eta_A}{4}},
\end{equation}
where $C_{\rm eq}(t)$ is the equilibrium correlation
function obtained when $t_w = 0$ (no shift at all) and $\eta_A$ and
$\eta_B$ are the usual critical exponents at temperatures $T_A$ and $T_B$.
The second and third term of the right hand side respectively 
account for the `transient' and the `memory'.
For $\ell_{T_A}(t) \ll \ell_{T_B}(t_w)$, 
the transient term is dominant and aging is observed 
through a power law decay.
When  $\ell_{T_A}(t) \gg \ell_{T_B}(t_w)$,
both terms combine to restore equilibrium, accounting for the memory.
Interestingly, the crossover time scale 
$t_c$ is given by 
\begin{equation}
\ell_{T_A}(t_c) \sim \ell_{T_B}(t_w). 
\label{cross}
\end{equation}
No separation of length scales amounts thus to $t_c \sim t_w$
as is observed in the electron glass.
Deviations from $t/t_w$, as observed at higher gate voltages $V$~\cite{ova1},
could be tentatively related via Eq.~(\ref{equation})
to a decrease of an activation energy at high $V$.
This effect could be more systematically investigated using 
Eqs.~(\ref{equation}) and (\ref{cross}).

\subsection{Colloidal suspension}

\begin{figure}[t]
\begin{center}
\psfig{file=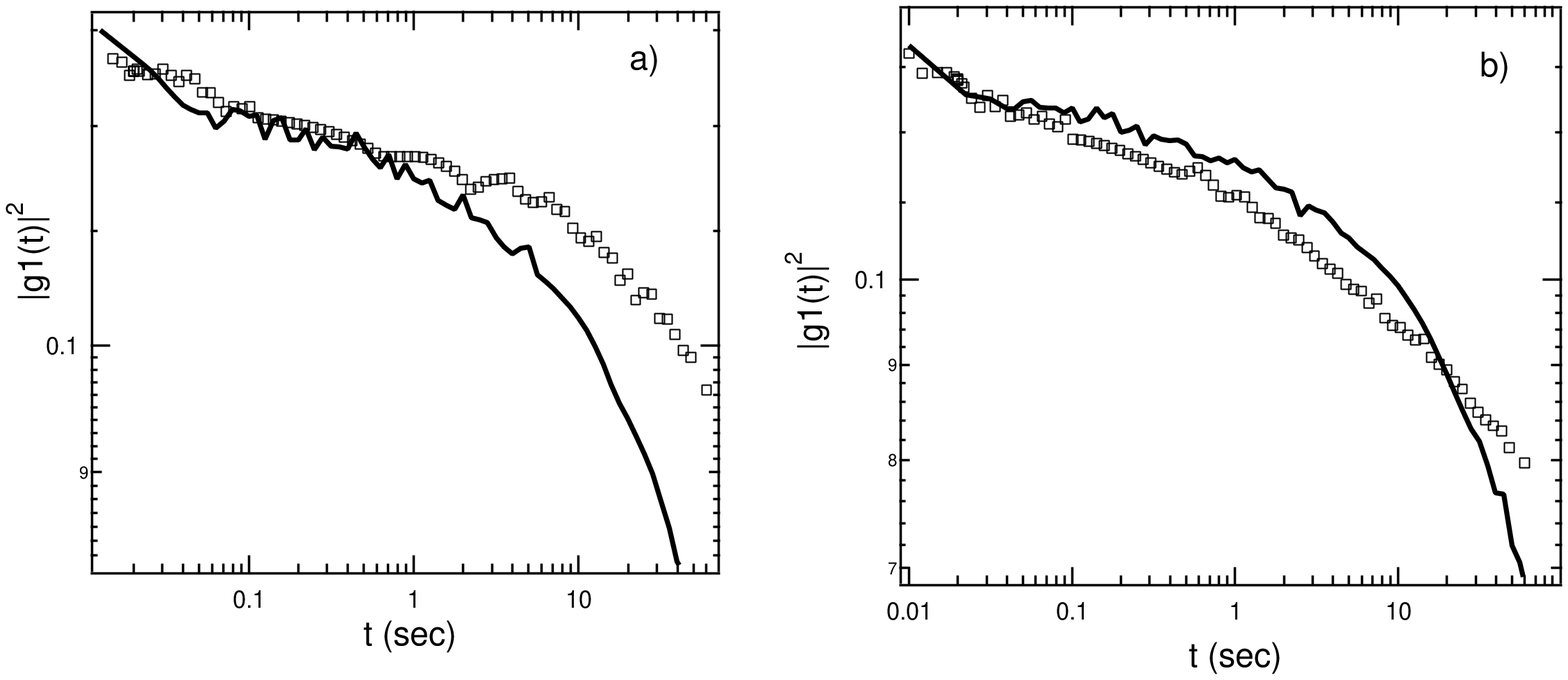,width=13cm}
\end{center}
\caption{Overaging in a colloidal suspension.
The data show two-time intensity correlation functions 
obtained with a multiple scattering technique.
Full lines refer to a simple aging experiment, 
squares to a `strain-shift' experiment with an intermediate 
strain of 5.9\%.
(a) $w=163$s. Overaging is observed as in Fig.~\ref{effects2}.
(b) $w=100$s. The strain difference 
is strong enough to partially rejuvenate the 
system (square curve below the full line at short times while
overaging is still observed at long times. 
This is the `intermediate' shift described in the text
where both rejuvenation and overaging are mixed.}
\label{overaging}
\end{figure}

Overaging was introduced in section~\ref{phenomeno}.
It is reproduced in Fig.~\ref{overaging} showing
a multiple light scattering experiment on
a colloidal suspension~\cite{virgile}.
Two interesting points in this experiment are (1) the control parameter
is an external 
oscillatory shear strain, as opposed to temperature
in `standard' shift experiments;
(2) no obvious coherence length 
is known to grow during the aging of colloids.
Therefore, such experiments could possibly be used to
characterize length scales in colloidal glasses.

In this context, `simple aging experiments' consist
in a `quench' from a shear strain of very large amplitude 
(typically more than 20 \%) mimicking a `high temperature'.
Standard $t/w$ scalings have been observed
in many similar soft glassy 
materials~\cite{paste,luca,lapo,virgile}.
It is thus natural 
to start and investigate more complex protocols, in the spirit
of shifts and cycles~\cite{virgile,virgile3,ozon}.
A first 
possibility is to quench first towards a small, but finite, shear 
strain during a time $t_w$, after which the shear is completely stopped.
One can then compare the results with a temperature shift experiment, 
$T_E \to T_A \to T_B < T_A$, which are more  
commonly performed.

In spin glasses, in agreement with the phenomenology 
developed above, the following results are known.
We already discussed the case of large $\Delta T = T_A - T_B$:
rejuvenation is observed~\cite{cycle1}.
For very small $\Delta T$, the relaxation of the system
after the shift has the same shape as in
a direct quench to $T_B$, but with an 
`effective age' $t_w^{\rm eff} > t_w$~\cite{cycle2}. 
Indeed, a small $\Delta T$ means that
the objects growing at $T_A$ will be almost unchanged
at $T_B$ (no rejuvenation).
However, due the property (iii), 
the time spent at higher temperature 
is more efficient in growing the coherence length. 
Hence, the relaxation is slower after the shift, the
system looking `older' or `overaged'.
This is observed in the colloidal suspension 
in a `strain-shift' experiment, see Fig.~\ref{overaging}(a).
More quantitatively, 
the effective age $t_w^{\rm eff}$ can be estimated
in terms of the coherence length as~\cite{BB,JPPRB,yosh} 
\begin{equation}
\ell(t_w^{\rm eff},T_B) \sim \ell(t_w,T_A).
\label{tweff}
\end{equation}
Repeating this protocol for various 
($t_w,T_A,T_B)$ gives thus 
quantitative informations on the growth law 
of the coherence length~\cite{BB,JPPRB,dupuis}.
Furthermore, Eq.~(\ref{equation})
shows that this protocol allows a
quantitative study of the energy
barriers encountered by the system 
during its nonequilibrium dynamics~\cite{JPPRB}. 

For intermediate $\Delta T$, of course, the result will
be a combination of rejuvenation and overaging. This has been observed in 
spin glasses~\cite{cycle2,BB,yosh} 
and in polymers~\cite{virgile2}.
That this is the case in the colloidal suspension with oscillatory
strain as a control parameter is demonstrated in 
Fig.~\ref{overaging}(b).

Again, systematic studies using Eqs.~(\ref{equation}) and (\ref{tweff}) could 
possibly lead to a quantitative characterization 
of a coherence length in this and other systems.

\section{Conclusion}

This paper is the result of a discussion group organized 
during this summer school. It
has become a kind of a review of 
what glassy dynamics looks like, as encountered 
in experiments performed on different materials using
diverse control parameters.
Although first observed in spin and structural glasses, it now is 
clear that these spectacular effects are quite generic. 
This points towards the necessity of having a simple 
interpretation of these phenomena and 
we have used the concept of a coherence length 
to develop one.
Note that  all the theoretical interpretations discussed 
in this paper rely on a 
some sort of hierarchical picture: hierarchy of traps levels, of time
scales in infinite-range glass models, 
of length scales
in spatial approaches. We have 
focused on 
the latter
because we think it gives 
greater physical insight.
We note, however, that 
definition, characterization and measurement 
of a coherence length is still 
largely an open
problem for most of the systems discussed here!
In this respect, we mentioned several times
that the experimental protocols discussed
in this paper and their interpretation, for instance Eq.~(\ref{cross}) 
and Eq.~(\ref{tweff}), constitute 
a promising starting point 
for progress on the crucial problem of length scales in glassy materials.

\section*{Acknowledgments}

We would like to express our gratitioude to the organizers for giving us the 
opportunity to appear in these proceedings.
We also acknowledge discussions and collaborations with 
G. Biroli, J.-P. Bouchaud, D. Fisher, P. Holdsworth, R. da Silveira,
H. Yoshino, and P. Young.


\begin{thebibliography}{99}
\bibitem{hike} Although `surfing on glassy phases' would have been
a possible title for this paper, we believe that `hiking' is better suited
in the context of a summer school taking place 
so close to the Alps and the `Mer de Glace'.
\end{thebibliography}
\end{document}